\documentstyle[aps,preprint,psfig,epsf]{revtex}
\begin{document}
\draft
\title{Properties of a magnetic superconductor with weak magnetization
- application to $ErNi_2B_2C$}
\author{Tai-Kai Ng and Wai-Tak Leung}
\address{Dept. of Physics, HKUST, Kowloon, Hong Kong}
\date{ \today }
\maketitle
\begin{abstract}
Using a Ginsburg-Landau free energy functional, we study the $H-T$
phase diagram of a weak magnetic superconductor, where the magnetization
from the magnetic component is marginal in supporting a spontaneous 
vortex phase in absence of external magnetic field. In particular, the
competition between the spiral state and spontaneous vortex phase is 
analysed. Our theory is applied to understand the magnetic properties 
of $ErNi_2B_2C$.
\end{abstract} 


\narrowtext

   Recently, there has been a renewed interest in the study of magnetic
superconductors\cite{nv1,nv2,ktyk,gam}, following the discoveries of the
magnetic superconductor $ErNi_2B_2C$, which, in the superconducting phase, 
has a vortex line-lattice with very unusal properties\cite{nv2}.  
Ng and Varma\cite{nv3} proposed that many of the unusual properties of
$ErNi_2B_2C$ can be understood by assuming that the system has an 
instability towards forming a {\em spontaneous vortex phase} at low 
temperatures below $T_{WF}\sim2.3K$. The coexistence of weak 
ferromagnetism and superconductivity in the system in the presence of 
external magnetic field was confirmed in neutron diffraction
measurement\cite{ktyk} and in magnetic hysterisis 
measurement\cite{gam} recently.

   The studies of Ng and Varma\cite{nv3} focused at the temperature 
range $T>T_{WF}$, when long-ranged magnetic ordering is absent. In this 
paper we shall study numerically the $H-T$ phase diagram of a weak 
magnetic superconductor, where the magnetization from the magnetic moment 
is not strong enough to support a spontaneous vortex phase in absence of
magnetic field. The behaviour of the system at both $T>T_{WF}$ and
$T<T_{WF}$ will be studied. In particular, we shall investigate the 
competition between the spiral phase\cite{t1,t2,t3} and spontaneous 
vortex phase. 

   We start with the Ginsburg-Landau Gibb's free energy functional 
{\em G} describing a magnetic superconductor system\cite{t2} in the 
presence of external field $\vec{H}$,
\begin{eqnarray}
\label{1}
{\em G} & = & \int{d^3x}[{1\over2}a|\psi|^2+{1\over4}b|\psi|^4+{\hbar^2\over
2m}|(\nabla-i{2e\over\hbar{c}}\vec{A})\psi|^2 + {\vec{B}^2\over8\pi}
\nonumber \\
 &  & +  
{1\over2}\alpha|\vec{M}|^2+{1\over4}\beta|\vec{M}|^4+{1\over2}\gamma^2
|\nabla\vec{M}|^2-\vec{B}.\vec{M}-{\vec{B}.\vec{H}\over4\pi}],
\end{eqnarray}
where $\vec{B}=\nabla\times\vec{A}$ is the total magnetic field, $\vec{M}$ 
is magnetization and $\psi$ is the superconducting order parameter. We 
assume here that the superconducting component is isotropic in space but 
the magnetic component is restricted to lie only on the $x-z$ plane, as is 
believed to be the case of the $ErNi_2B_2C$ compound\cite{nv1}. The 
anisotropy within the $x-z$ plane is neglected in this study. The effect 
of in-plane anisotropy will be discussed at the end of the paper. We shall 
work in the London limit where contributions from vortex cores will be
neglected, and shall consider only fluctuations in the {\em phase} of
the superconducting order parameter $\psi$. In this limit, we obtain an
effective GL functional $G_{eff}$ in terms of the magnetization 
component and the vorticity (vortex density) field $\vec{\kappa}(\vec{x})$\cite{nvlong}, where
\begin{mathletters}
\label{geff}
\begin{eqnarray}
\label{geff1}
{\em G_{eff}}[\vec{M},\vec{\kappa}] & = & \int{d}^3x\{-{a\over4}\rho
+{1\over8\pi}(\vec{\kappa})^2-{1\over4\pi}\vec{\kappa}.(
\vec{H}+4\pi\vec{M})+{1\over2}\alpha|\vec{M}|^2
+{1\over4}\beta|\vec{M}|^4+{1\over2}\gamma|\nabla\vec{M}|^2\}
\\ \nonumber
& & -{1\over4}\int{d}^3xd^3x'(\kappa-H-4\pi{M})_{\mu}(\vec{x})
(\Pi^{-1})_{\mu\nu}(\vec{x}-\vec{x}')(\kappa-H-4\pi{M})_{\nu}(\vec{x}'),
\end{eqnarray}
with fourier transform of $(\Pi^{-1})_{\mu\nu}(\vec{x})$ given by
\begin{equation}
\label{pi-1}
(\Pi^{-1})_{\mu\nu}(\vec{q})={1\over2\pi}\left(
{\lambda_0^2q^2\over(1+\lambda_0^2q^2)}
[\delta_{\mu\nu}-{q_{\mu}q_{\nu}\over{q}^2}]\right).
\end{equation}
\end{mathletters}
$\rho=|\psi|^2$ and $\lambda_0$ are the superfluid density and London
penetration depth, respectively, in GL theory in the absence of the 
magnetic component. Notice that only transverse magnetization
($\nabla\times\vec{M}\neq0$) couples to vortices. We shall study the 
spontaneous vortex phase and spiral phase of the system in this paper 
using approximate, variational solutions of $G_{eff}$. We first 
consider the case when the external magnetic field $H$ is along $z$-axis, 
where the magnetization couples directly to $H$.

   We consider trial solutions of $G_{eff}$ of form
\begin{eqnarray}
\label{trial}
\vec{M}(\vec{r},z) & = & M_o\hat{z}+M_1cos(Qy)\hat{z}+M_2sin(Qy)\hat{x},
\\  \nonumber
\kappa(\vec{r},z) & = & \Phi_o\sum_i\hat{z}\delta^{(2)}(\vec{r}-\vec{R}_i),
\end{eqnarray}
where $\vec{r}=(x,y)$. The solution represents an ideal flux-line lattice 
in $\hat{z}$ direction with straight vortex lines, where $\vec{R}_i=(X_i,Y_i)$
represents the position of vortices in the $x-y$ plane which form a 
regular lattice. The magnetic component may form a spiral state with 
magnetization lying on the $x-z$ plane, where $\vec{Q}=Q\hat{y}$ is a 
spiral wave vector yet to be determined, or form a spontaneous vortex
phase with $M_o\neq0$. Notice that we have also included the
possibilities of co-existence phases ($M_o,M_1,M_2\neq0$)in our 
trial solution. 
 
 Substituting Eq.\ (\ref{trial}) into Eq.\ (\ref{geff}) we obtain after
some algebra
\begin{eqnarray}
\label{gvar}
{G\over{V}} & = & {\alpha\over2}(M_o^2+{(M_1^2+M_2^2)\over2})+{\beta\over4}
\left(M_o^4+M_o^2{(M_1^2+M_2^2)\over2}+2M_o^2M_1^2+{(M_1^4+M_2^4)\over2}
+{M_1^2M_2^2\over4}\right)  \\  \nonumber
& & +{\gamma\over2}Q^2{(M_1^2+M_2^2)\over2}-{1\over4\pi}B(H+4\pi{M}_o)+
{B^2\over8\pi}-{2\pi\lambda_0^2Q^2\over1+\lambda_0^2Q^2}{(M_1^2+M_2^2)\over2}
\\  \nonumber
& & +{1\over8\pi}\sum_{\vec{q}_N}{B^2\over(1+\lambda_0^2q_N^2)}-\delta_{\vec{Q},
\vec{q}_N}{BM_1\over(1+\lambda_0^2Q^2)},
\end{eqnarray}
where $V$ is the volume of the system. The fourier transform of vortex
density field $\vec{\kappa}(\vec{q},q_z)$ is 
\[
\vec{\kappa}^{(0)}(\vec{q},q_z)=\delta(q_z)\hat{z}B\delta^{(2)}
(\vec{q}-\vec{q}_N)V,
\]
where $\vec{q}_N$'s are reciprocal lattice vectors and $B=n\Phi_0$ is the
average magnetic field trapped in the flux-line lattice, $n$ is the density 
of vortex lines. Notice that $M_1$ couples to the flux-line lattice when
$\vec{Q}=\vec{q}_N$ for some $\vec{q}_N$, reflecting the fact that the 
existence of a flux-line lattice induces fluctuations in magnetization 
which is commensurate with the lattice.

   To simplify our calculation we replace the sum 
$S=\sum_{q_N}{B^2\over8\pi(1+\lambda_0^2q_N^2)}$ by the approximate value 
$S=BH_{c1}/4\pi$\cite{tinkham} which is valid at $H\sim{H}_{c1}$, 
where $H_{c1}$ is the lower critical field in the absence of the 
magnetic component. Logarithmic correction to $S$ arises at intermediate 
value of $H_{c2}>H>H_{c1}$ which we shall neglect in the following since 
we shall be interested mainly at 
qualitative features at regions around $H\sim{H}_{c1}$. Minimizing the 
energy with respect to $Q$ and assuming that $\vec{Q}\neq\vec{q}_N$ we 
obtain $Q^{-2}=Q_s^{-2}\sim\delta^{1\over2}\lambda_0^2$, where $Q_s$ is 
the wave vector characterizing spiral instability\cite{t1,t2} in 
magnetic superconductors, $\delta=\gamma/4\pi\lambda_0^2$. To examine
the importance of magnetization fluctuations commensurate with the 
flux-line lattice, we have also consider 
$Q=q_1\sim(Bln(\kappa)/H_{c1})^{1\over2}\lambda_0^{-1}$ in our study, 
where $q_1$ is the smallest reciprocal lattice vector of the flux-line 
lattice. The corresponding free energies are minimized with respect to
$B,M_o,M_1,M_2$ at various temperatures and external magnetic fields $H$ 
to determine the phase diagram. We have chosen parameters such that
temperatures are measured relative to superconducting transition 
temperature $T_c$, which is set to be $T_c=10$. We have chosen also
$\kappa=\lambda_0/\xi=5$, and $\alpha=\alpha_o(T/t_m-1)$, where $\alpha_o=10$ 
and $\beta=20$. $t_m$ is chosen such that $T_{M}=3$ (or $0.3T_c$), where 
$T_{M}$ is the ferromagnetic transition temperature in the absence of 
superconducting component. Notice that in the presence of the superconducting 
component, the actual magnetic transition temperature $T_{WF}$ is below 
$T_{M}$ because the Meissner effect forbids appearance of
uniform magnetization in a superconductor\cite{t1,t2}. Moreover, 
the zero field magnetization $M_{sat}$ at $T=0$ in absence of 
superconducting component is chosen to be $M_{sat}=H_{c1}/4\pi$, i.e. the 
system is marginal in supporting a spontaneous vortex phase.  

  Minimizing the energy of the system we find that magnetic fluctuations
commensurate with flux-line lattice is always unimportant compared with
the spiral instability and the system is in a spiral state with 
$M_1={M}_2$ and $M_o=0$ at zero magnetic field at temperatures
$T<T_{WF}\sim2.74$. The system goes through a second order phase transition 
to a non-magnetic superconducting state at $T=T_{WF}$. At $T<T_{WF}$, the 
system stays at the spiral state when external magnetic field is applied, 
until $H\rightarrow{H}_{cs}<H_{c1}$, where the system undergoes a 
{\em first order} phase transition into a magnetic-field assisted spontaneous 
vortex phase with finite $M_o$, and with $M_1=0$ and small $M_2$. The 
presence of $M_2\neq0$ is a residue effect of spiral instability in the 
spontaneous vortex state. We find numerically that $M_2$ reduces 
rapidly to zero as external field $H$ increases further. The $H-T$
phase diagram is shown in Fig.1, where $(VP), (SP), (MP)$ denotes the
magnetic field-assisted spontaneous vortex phase, spiral phase and Meissner
phase, respectively.  $H_{c1}^*$ is the zero temperature lower critical 
field in the absence of the magnetic component. The dotted and 
dashed lines give $H_{c2}$ and $H_{c1}$ of the superconducting 
component, respectively, in the absence of 
the magnetic component. The solid line gives $H_{cs}$ and represents a 
first order transition line from the spiral phase to the vortex phase at
$T<T_{WF}$, and represents the usual transition from Meissner phase to
vortex phase at $T>T_{WF}$. The dot-dashed line denotes the second order
transition from Meissner phase to spiral phase at $H<H_{cs}$.
\begin{figure}[htb]
\begin{center}
\leavevmode
\epsfxsize=4in
\epsfbox{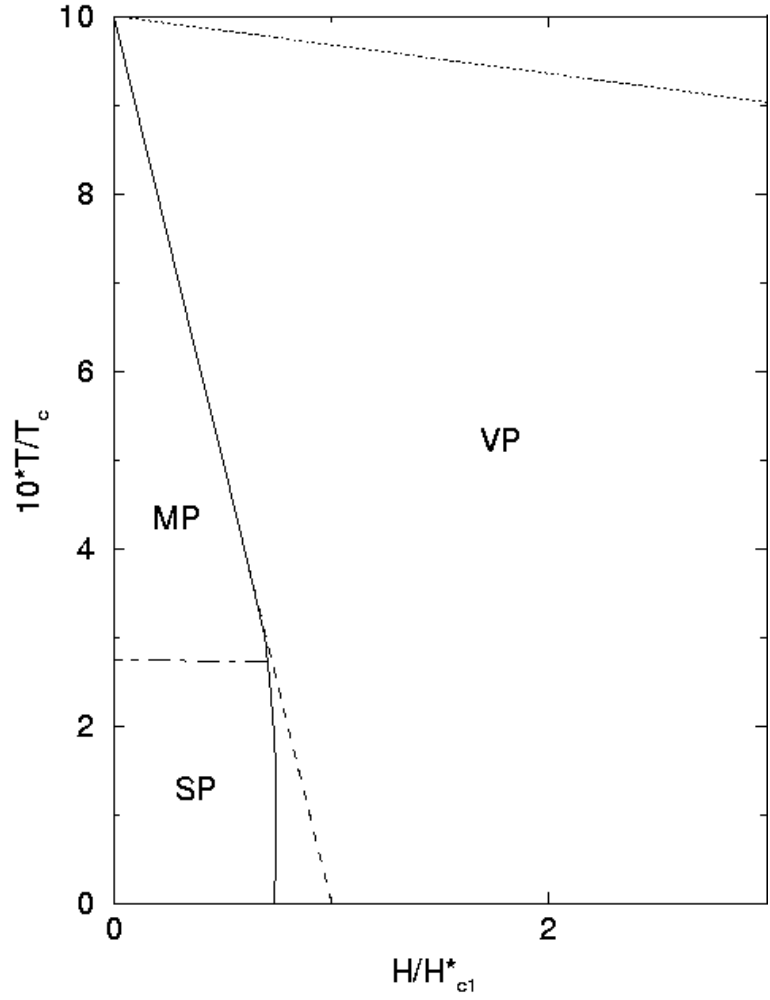}
\caption{H-T phase diagram of the magnetic superconductor}
\end{center}
\end{figure}

  To make comparisons with experiments we show in Fig.2 the total magnetization
$M=(B-H)/4\pi$ versus external magnetic field $H/H_{c1}^*$ at three 
different temperatures $T=0.0,2.0,4.0$,
\begin{figure}[htb]
\begin{center}
\leavevmode
\epsfxsize=4in
\epsfbox{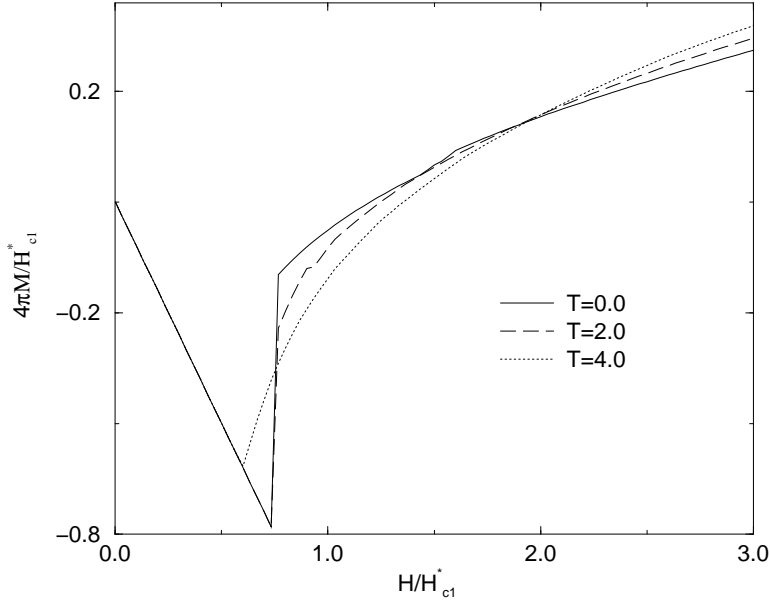}
\caption{Magnetization versus magnetic curve for at three different temperatures T=0.0, 2.0, 4.0.}
\end{center}
\end{figure}
The qualitative behaviour of the $M-H$ curve is similar to those observed
experimentally in the $ErNi_2B_2C$ compound at low temperature\cite{ktyk}.
Notice a sudden jump in magnetization at $H=H_{cs}$ appears 
when $T<T_{WF}$, signaling the first-order transition to magnetic field 
assisted spontaneous vortex phase. This sudden jump is absent at
$T\geq{T}_{WF}$, justifying the identification of $T_{WF}$ as the spiral 
state transition temperature. The sudden jump in magnetization will probably
be smeared out by disorder in realistic samples, and $H_{cs}$ can be 
considered as an effective lower critical field observed in magnetic
superconductors at $T<T_{WF}$. Notice that in previous analysis of Ng 
and Varma\cite{nv3}, the lower critical field $H_{c1}$ for magnetic
superconductors was computed from single vortex line energy and it was 
found that the value of $H_{c1}$ is not much affected by the magnetic 
component in the spiral state. However, we find here that $H_{cs}$ is 
smaller than $H_{c1}$ at $T<T_{WF}$, indicating that an effective 
reduction in $H_{c1}$ can occur because of the first order transition to 
the magnetic field assisted spontaneous vortex phase. As $M_{sat}$ 
increases further, we find numerically that $H_{cs}$ starts to decrease 
as temperature decreases until when $4\pi{M}_{sat}\sim4H_{c1}^*$ where
$H_{cs}\rightarrow0$ at $T\rightarrow0$, indicating a zero temperature 
phase transition of the system to the spontaneous vortex phase 
at zero magnetic field. As $M_{sat}$ increases further, the region where
the spontaneous vortex state is stable at zero field expands further and 
the spiral state becomes stable only at a finite temperature range 
$0<T_{sp}<T<T_{WF}$, in agreement with results from previous studies\cite{t2,t3}.

  Recently it was observed in magnetization studies of the $ErNi_2B_2C$
compound that substantial hystersis developed in the $M-H$ loop at
temperature $T\leq{T}_{WF}$\cite{gam}, and the result was interpreted as
coming from a $\sim$ three fold increase of the pinning force of the
flux line lattice at $T<T_{WF}$. Here we shall consider the effect of
spiral phase in magnetic hystersis. We note that the spiral phase and
spontaneous vortex phase are locally stable phases at $T\leq{T}_{WF}$ over 
a wide range of magnetic fields. The spiral phase is (locally) stable at
$H<H_1^l$, and the spontaneous vortex state is (locally) stable at 
$H>H_2^l$, where $H_1^l(H_2^l)$ depends on the parameters in the
GL functional, and is a decreasing (increasing) function of
temperature, with $H_1^l>H_2^l$ in general. We propose that the huge 
magnetic hystersis observed at $T<T_{WF}$ is a result of the system being
trapped in the two meta-stable states when magnetic field increases
(decreases) at temperature $T<T_{WF}$. When magnetic field increases from 
zero, the system is trapped in the spiral state until $H\geq{H}_1^l$, 
where the spiral state becomes locally unstable and the system goes to 
the spontaneous vortex state. Similarly the system is trapped in the 
spontaneous vortex state when magnetic field decreases, until 
$H\leq{H}_2^l$, where the system goes back to the spiral state. As a 
result, huge magnetic hystersis can be observed at magnetic field range $H_2^l<H<H_1^l$. 
\begin{figure}[htb]
\begin{center}
\leavevmode
\epsfxsize=4in
\epsfbox{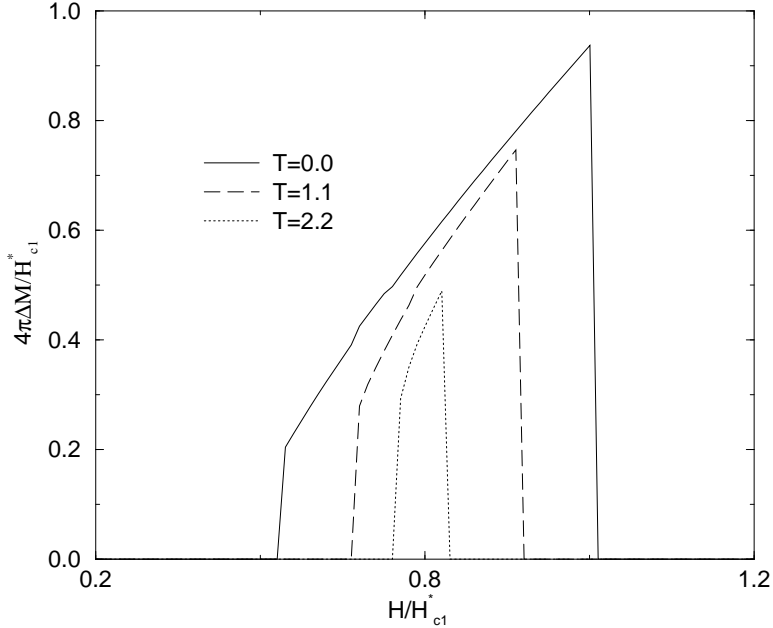}
\caption{Magnetic hystersis between spiral and (magnetic-field assisted) spontaneous vortex phase at three different temperatures T=0.0, 1.1, 2.2.}
\end{center}
\end{figure}
In figure 3 we show the hystersis magnetization $\Delta{M}=
|M_{up}-M_{down}|$ as function of external magnetic field $H/H_{c1}^*$ 
at three different temperatures $T=0.0,1.1,2.2$, assuming the above picture. 
The narrowing of the hystersis region and reduction in $\Delta{M}$ as 
temperature increases is in qualitative agreement with what is observed
experimentally\cite{gam}. Notice that first order jumps of $\Delta{M}$ 
is found at $H=H_1^l$ and $H=H_2^l$ in our theory. These jumps should be 
smeared our by defects in real samples. Notice also that additional 
features are observed experimentally at $H\sim{H}_l^1$ at low
temperatures\cite{gam}, suggesting that there exist additional 
structures in the spiral-spontaneous-vortex state phase diagram not 
included in our simple variational solution.
 
  Next we consider the case when the applied magnetic field is almost
perpendicular to the magnetic plane and makes a small angle $\theta$ 
with the $\hat{y}$-axis. In this case, the coupling 
between magnetization and external field is weak. However, it was observed 
experimentally that the flux-line lattice seen to rotate away from the applied 
field direction at temperature $T\rightarrow{T}_{WF}$\cite{nv2}. This
observation can be explained qualitatively by development of in-plane
magnetization as $T\rightarrow{T}_{WF}$\cite{nv3}. Here we shall study
numerically this tilting effect at temperatures both above and below 
$T_{WF}$. 

  The trial solutions of $G_{eff}$ we consider is similar to Eq.(3),
except that $\kappa(\vec{r},z)$ is in Eq.(3) is replaced by
\[
\kappa(\vec{r},z)=\Phi_o\sum_i(cos(\theta_t)\hat{y}+sin(\theta_t)\hat{z})
\delta^{(2)}(\vec{r}-\vec{R}_i), \] 
where $\theta_t$ is the angle the flux-line lattice made with the
$\hat{y}$-axis. Correspondingly, the term $-B(H+4\pi{M}_o)/4\pi$ in Eq.(4)
is replaced by
\begin{equation}
\label{tilt}
-{B(H+4\pi{M}_o)\over4\pi}\rightarrow{-1\over4\pi}
B\left(Hcos(\theta-\theta_t)+4\pi{M}_osin(\theta_t)\right). 
\end{equation}
   Minimizing the energy of the system we obtain $tan(\theta_t)=
(Hsin(\theta)+4\pi{M}_o)/Hcos(\theta)$. 
\begin{figure}[htb]
\begin{center}
\leavevmode
\epsfxsize=4in
\epsfbox{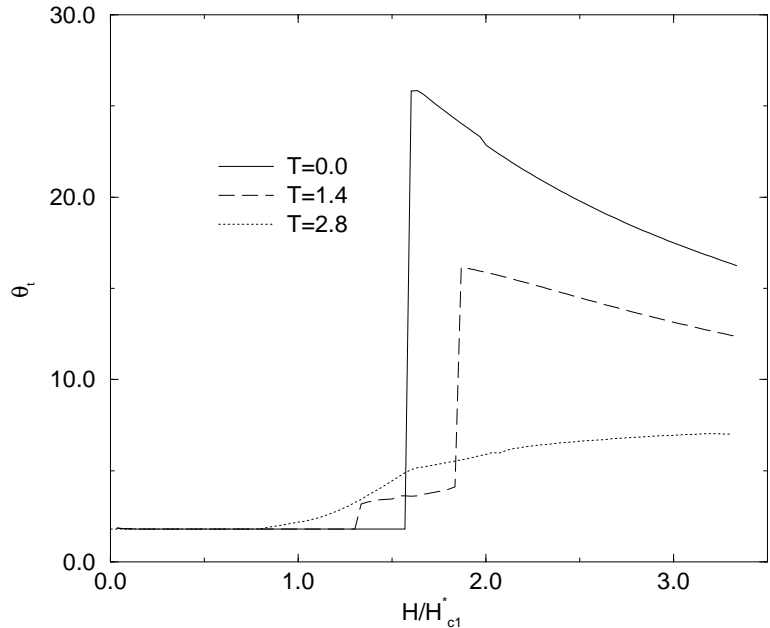}
\caption{Angle of the flux-line lattice with the $\hat{y}$-axia as a function of magnetic field at three different temperatures $T=0.0,1.4,2.4$. The angle between applied magnetic field and $\hat{y}$-axia is $1.8^o$.}
\end{center}
\end{figure}
The resulting angle $\theta_t$ is shown in figure 4 as a function of 
external magnetic field $H/H_{c1}^*$ at three different temperatures
$T=0.0,1.4,2.8$ with $\theta=0.01\pi(1.8^o)$. At $T=0$, $\theta_t$ 
remains at $1.8^o$ at low field, indicating that the system is in the 
spiral state. Flux-line lattice with $\theta_t=\theta$ ($M_o=0$) begins 
to form at $H\sim{H}_{c1}$, because the coupling of the magnetization to 
the applied field is very weak at small $\theta$. 
A first order phase transition to the (magnetic-field assisted)
spontaneous vortex phase occurs at $H\sim1.6H_{c1}^*$, where in-plane
magnetization $M_o$ becomes nonzero ($M_1=0,M_2$ small) and $\theta_t$ 
jumps to a value $\sim25^o$. As magnetic field increases further, 
$\theta_t$ decreases because the in-plane magnetization is already 
near saturation and the rate of increase in $M_o$ is slower than the 
rate of increase in $H$. Similar result is also found at $T=1.4$, except 
that the first order transition occurs at a higher field and $\theta_t$ is smaller. We find numerically that there exists also a small region between the 
spiral and spontaneous vortex phase where the two states are co-existing 
in the system with a small value of $M_o$ but large $M_1\sim{M}_2$, 
with $\theta_t\sim4^o$. This state is stable only at intermediate 
temperatures below $T_{WF}$ and at a narrow range of magnetic field
$H>H_{c1}$. At temperatures above but close to $T_{WF}$, we find a 
continous increase of $\theta_t$ as function of $H$ at $H>H_{c1}$, 
in agreement with previous experimental observation\cite{nv2}. When 
$\theta=0$, direct first order transitions to spontaneously-tilted 
flux-line lattice phase\cite{nv3} is also found at 
$T<T_{WF}$, when magnetic field is large enough. 

   Summarizing, using simple variational solutions of the GL free energy
functional, we have investigated numerically the magnetic behaviours of a
magnetic superconductor where the magnetic component is weak and 
marginal in supporting a spontaneous vortex phase. We have studied both
the cases when magnetic field is parallel to the magnetic plane, and
when it is almost perpendicular to the magnetic plane. Our results are 
in general agreement with experimental observations. Furthermore, we
have made several theoretical predictions in this paper, including the
appearance of spiral phase at zero field at $T<T_{WF}$, which can be
tested by neutron scattering experiment. Notice that we have neglected
in-plane magnetic anisotropy in our study. In the case of $ErNi_2B_2C$ 
where magnetization are strongly confined to point at either $\hat{x}$
or $\hat{z}$ directions, a smooth spiral phase is impossible and will 
be replaced by sharp domain structures\cite{nv3} that can be disordered 
easily. In this case the sharp elastic neutron peaks at spiral 
wavevector $\vec{Q}_s$ will be replaced by a broadened peak centered 
at wavevector $\vec{q}=0$ with width of order $Q_s$. We predict also 
a first order transition to spontaneous vortex phase when external 
field is strong
enough, both when the field direction is in the magnetic plane, and
when it is almost perpendicular to the plane. In general a stronger 
magnetic field is needed for this transition to occur when the angle 
between the applied field and the magnetic plane is larger. When the
magnetic field is almost perpendicular to the plane, the first order
transition to spontaneous vortex phase appears as a spontaneous
tilting transition of the flux-line lattice, which can be tested
experimentally. 

  T.K. Ng acknowlegdes many helpful discussions with C.M.
Varma. This work is supported by Hong Kong UGC through Grant No.
HKUST6124/98P.




\begin{references}
\bibitem{nv1} B. K. Cho, et al., Phys. Rev. B{\bf 52}, 3684
(1995); P. C. Canfield, S. L. Budko and B. K. Cho, Physica
(Amsterdam) {\bf 262C}, 249 (1996); P. Dervengas, et al., Phys.
Rev. B{\bf 53}, 8506 (1996).
\bibitem{nv2} U. Yaron, et al., Nature {\bf 82}, 236 (1996).
\bibitem{ktyk}J. Kawano {\em et.al.}, J. Phys. Chem. Solids {\bf 60}, 
 1053 (1999).
\bibitem{gam} P.L. Gammel {\em et.al.}, \prl {\bf 84}, 2497 (2000).
\bibitem{nv3} T. K. Ng and C. M. Varma, \prl {\bf 78},
339 (1997); T.K. Ng and C.M. Varma, \prl{\bf 78}, 3745 (1997).
\bibitem{t1} H.S. Greenside, E.I. Blount and C.M. Varma, \prl{\bf 46}, 49 (1981)
\bibitem{t2} E.I. Blount and C.M. Varma, \prl{\bf 42}, 1079 (1979).
\bibitem{t3} M. Tachiki {\em et.al.}, Sol. State Comm.{\bf 31}, 927 (1979);
     {\em ibid}{\bf 34}, 19 (1980).
\bibitem{nvlong} T.K. Ng and C.M. Varma, \prb {\bf 58}, 11624 (1998).
\bibitem{tinkham} M. Tinkham, {\em Introduction to Superconductivity}
 (Krieger, Malabar, FL, 1980).
\end{references}
\end{document}